\documentclass[useAMS,usenatbib]{mn2e}

\usepackage{graphicx}
\usepackage{natbib}

\title[The binary population in globular clusters]{The evolution of the binary population in
globular clusters: a full analytical computation}
\author[Sollima]{A. Sollima$^{1}$\thanks{E-mail:
asollima@iac.es (AS)}\\
$^{1}$Istituto de Astrofisica de Canarias, C/via Lactea s/n, San Cristobal de La
Laguna, Tenerife, 38205-E, Spain}
\begin{document}

\date{Accepted 2008 August ??; Received 2008 August ??; in original form
2008 July ??}

\pagerange{\pageref{firstpage}--\pageref{lastpage}} \pubyear{2008}

\maketitle

\label{firstpage}

\begin{abstract}
I present a simplified analytical model that simulates the evolution of the binary population in
a dynamically evolving globular cluster.
A number of simulations have been run spanning a wide range in initial cluster
and environmental conditions by taking into account the main mechanisms of
formation and destruction of binary systems.
Following this approach, I investigate the evolution of the fraction, 
the radial distribution, the distribution of mass ratios and periods of the
binary population.
According to these simulations, the fraction of surviving binaries appears to be 
dominated by the processes of binary ionization and evaporation. In particular, the frequency of
binary systems changes by a factor 1-5 depending on the initial conditions and
on the assumed initial distribution of periods.
The comparison with the existing estimates of binary fractions in Galactic
globular clusters suggests that significant variations in the initial binary
content could exist among the analysed globular cluster.
This model has been also used to explain the observed discrepancy found
between the most recent N-body and Monte Carlo simulations in the literature.

\end{abstract}

\begin{keywords}
stellar dynamics -- methods: analytical -- 
stars: binaries: general -- Galaxy: globular clusters: general
\end{keywords}

\section{Introduction} 
\label{intro}

The study of the evolution of binary stars both in the solar neighborood and in
stellar clusters is one of the most interesting topic of stellar astrophysics.
Binarity, under particular conditions, induces the onset of nuclear reactions
leading to the formation of peculiar objects like novae and determines the fate 
of low-mass stars leading to SN~Ia explosions.
In collisional systems binaries provide the gravitational fuel that can delay
and eventually stop and reverse the process of core collapse in globular clusters (see
Hut et al. 1992 and references therein). 
Furthermore, the evolution of binaries in star clusters can produce peculiar
stellar objects of
astrophysic interest like blue stragglers, cataclysmic variables, low-mass X-ray
binaries, millisecond pulsars, etc. (see Bailyn 1995 and reference therein). 

From an observational point of view, there is an extensive literature on the
analysis of the main characteristics of the binary population of the Galactic 
field (Duquennoy \& Mayor 1991; Halbwachs et al. 2003 and references therein).
Conversely, the analysis of the binary population in star clusters is
still limited to small samples of open clusters (Bica \& Bonatto 2005) and  
to few globular clusters (Yan \& Mateo 1984; Romani \& Weinberg 1991; Bolte 1992; 
Yan \& Cohen 1996; Yan \& Reid 1996;
Rubenstein \& Bailyn 1997; Albrow et al. 2001; Bellazzini et al. 2002; 
Clark, Sandquist \& Bolte 2004; Zhao \& Bailyn 2005).
More recently, Sollima et al. (2007) estimated the binary fraction in 13 low-density
Galactic globular clusters. They found that the fractions of binary systems
span a wide range comprised between 10-50\%. This spread could be due either to 
differences in the primordial binary fractions or to a different
evolution of the binary populations.  

Theoretical studies addressed to the study of the evolution of the properties of
the binary population in globular clusters are mainly based on two different
approaches: {\it i)} full N-body simulations and {\it ii)} Monte Carlo
simulations. The former follows the dynamical evolution of the stellar system 
assuming simplified treatments of single and binary star evolution (see
Portegies Zwart et al. 2001; Shara \& Hurley 2002; Hurley \& Shara 2003; 
Trenti, Heggie \& Hut 2007; Hurley et al. 2007).
This approach is extremely expensive computationally and have been often 
performed with unrealistically small numbers of binaries.
The latter method uses a binary population synthesis code to evolve large 
numbers of stars and binaries by introducing a simple treatment of dynamics. 
In this type of approach it is often assumed that all the relevant parameters 
of the cluster (central density, velocity dispersion, mass, etc.) remain 
constant during the cluster evolution (see Hut, McMillan \& Romani 1992;
Di Stefano \& Rappaport 1994; Sigurdsson \& Phinney 1995; Portegies Zwart et al.
1997; Portegies Zwart, Hut \& Verbunt 1997; Davies 1995; Davies \& Benz 1995; Davies 1997; Rasio et al. 2000;
Smith \& Bonnell 2001; Ivanova et al. 2005).
The most recent works performed following these two different approaches leaded 
to apparently contradictory results: while Monte Carlo simulations (Ivanova et
al. 2005) predict a strong depletion of the binary population in the cluster 
core, the N-body simulations performed by Hurley et al. 2007 show a progressive
increase of the core binary fraction as a function of time.

In this paper I study the evolution 
of the binary population in a dynamically evolving globular cluster using a
simplified analytical approach. Following this approach, I investigate the evolution of 
the fraction, the radial distribution, the distribution of mass ratios and 
periods of the binary population. A number of simulations have been run 
spanning a wide range in initial cluster and environmental conditions.

In Sect. \ref{code} I describe in detail the code. In Sect. \ref{test} 
I present the complete set of
simulations and discuss the dependence of the results on the underlying
assumptions. 
In Sect. \ref{res} the predicted evolution of 
the binary fraction as a function of the initial cluster structural and
environmental parameters is presented. Sect. \ref{rad} is
devoted to the study of the radial distribution of binary systems. 
In Sect. \ref{per} and \ref{fm} the distribution of periods and mass ratios are investigated
together with their dependence on the environmental conditions. A comparison
with the most recent N-body and Monte Carlo simulations is performed in Sect.
\ref{confteo}.
In Sect. \ref{confobs} the predictions of the code are compared with the
estimates of binary fractions in globular clusters available in the literature.   
Finally, I discuss the obtained results in Sect. \ref{concl}.     

For clearity, I list below the notation used throughout the
paper.\\

\begin{tabular}{@{}p{2.5cm}l@{}}
{\it General} &\\
$\rho_{r}$     & density at the distance $r$ from the \\
               & cluster center\\
$\sigma_{v,r}$ & velocity dispersion at the distance\\
               & $r$\\
$r_{c}$        & core radius\\
$r_{t}$        & tidal radius\\
$\xi$          & binary fraction\\
$\xi_{c}$      & core binary fraction\\
$\nu$          & fraction of evaporating systems\\
$t_{9}$        & cluster age\\
$\alpha$       & IMF power-law index\\
$<m>$          & average object mass \\
               & (binaries+single stars)\\
$N_{sys}$      & number of cluster objects\\
               & (binaries+single stars)\\
$M$            & cluster mass\\
$M_{c}$        & cluster core mass\\	       
$t_{r}$        & local relaxation time at the\\
               & distance $r$\\
$r_{rel}$      & relaxation radius\\	       
$x$            & polytropic index\\
\end{tabular}

\begin{tabular}{@{}p{2.5cm}l@{}}
$f(v,r)$       & distribution of velocities at the\\
               & distance $r$\\
$\sigma_{X}$   & cross section of the process X\\
$v_{e,r}$      & escape velocity at the distance $r$\\
$G$            & Newton constant\\
$log~\Lambda$  & Coulomb logarithm\\
$\Phi_{r}$     & potential at the distance $r$\\
$\epsilon_{r}$ & energy per unity of mass at the\\
               & distance $r$\\
\end{tabular}

\begin{tabular}{@{}p{2.5cm}l@{}}
{\it Binaries} &\\
$m_{1}$                      & mass of the primary star\\
$m_{2}$                      & mass of the secondary star\\
$P$                          & period\\
$\eta$                       & hardness parameter\\
$E_{b}$                      & binding energy\\
$N_{b}(m_{1},m_{2},log~P)$   & number of binaries with\\
                             & parameters $m_{1},m_{2},log~P$\\
$n_{b,r}(m_{1},m_{2},log~P)$ & number density of binaries with\\
                             & parameters $m_{1},m_{2},log~P$ at the\\
                             & distance $r$ \\
$g(log~P)$                   & distribution of periods\\			   
$q$                          & mass ratio\\
\end{tabular}

\begin{tabular}{@{}p{2.5cm}l@{}}
{\it Single stars} &\\
$m_{s}$          & mass\\
$N_{s}(m_{s})$   & Number of single stars with mass\\
                 & $m_{s}$\\
$n_{s,r}(m_{s})$ & number density of single stars\\
                 & with mass $m_{s}$ at the distance $r$\\
\end{tabular}

\section{Method} 
\label{code}

Each simulation is characterized by four initial parameters: the cluster central
density $\rho_{0}$, the central velocity dispersion $\sigma_{v,0}$, the initial
binary fraction $\xi$ and the rate of evaporating systems $\nu$.

The code simulates the evolution of the number of binaries and single
stars by taking into account the main processes of
formation, destruction and evolution of binaries and single stars.
In particular, the following processes have been considered:

\begin{itemize}
\item{tidal capture}
\item{direct collisions}
\item{exchanges}
\item{collisional hardening}
\item{binaries ionization}
\item{stellar evolution}
\item{coalescence}
\item{mass segregation}
\item{evaporation}
\end{itemize}

A detailed description of the analytical treatment of each of these processes is
given in the sub-sections from \ref{tid} to \ref{evap}.
Moreover, during its evolution, the cluster dynamical parameters
are assumed to evolve (see Sect. \ref{clust}). 

The binary population of the system has been divided in several groups 
according to the masses of the primary and secondary stars ($m_{1}$ and $m_{2}$), 
and to the period ($P$).
Similarly, the population of single stars has been divided in groups of masses
($m_{s}$). 

The evolution of each population of single and binary stars is calculated in
layers of variable width located at different distances from the cluster center.

At the beginning of the simulation, the density and the distribution of 
velocities of both single and binary stars is assumed to follow the radial 
behaviour of a mono-mass King profile, as expected for a non-relaxed stellar 
system. 
During this initial stage, the cluster is assumed to be a non-collisional 
system. Therefore, the only processes at work are those related to stellar 
evolution and evaporation.

After this initial stage, the evolution of the cluster and its binary population
is followed in time-steps.
For each time-step the adopted general procedure is schematically the
following: 

\begin{enumerate}
\item{The code calculates the maximum radius in which 
the local relaxation time is smaller than the cluster age
($relaxation~radius$; $r_{rel}$). The local relaxation time at the radius $r$ has been 
calculated using the relation
\begin{equation}
\label{eq_tr}
t_{r}=0.34 \frac{\sigma_{v,r}^3}{G^2 \rho_{r} <m> log \Lambda} ~~\mbox{Binney \&
Tremaine (1987)}
\end{equation}
with $\Lambda=0.4 N_{sys}$.} 
\item{The number of newly formed and destroyed binary systems 
have been calculated in all the layers located inside the 
$relaxation~radius$. The number of binaries $N_{b}(m_{1},m_{2},log~P)$ and single 
stars $N_{s}(m_{s})$ belonging to each group have been consequently updated;}
\item{A set of partially-relaxed multi-mass King profiles has been calculated (see
Appendix A). The density and velocity distribution of each
sub-population of binary and single stars have been assumed to follow a given 
profile according to their systemic masses;}
\item{The mass of the cluster has been calculated and the values of $\rho_{0}$,
$\sigma_{v,0}$ and $W_{0}$ have been updated (see Sect. \ref{clust});}
\end{enumerate}

Point {\it i) to {\it iv)}} have been repeated until the cluster reach an
age of 13 Gyr.
In the next sub-section I will report the assumptions made in all the
simulations presented in this paper.

\subsection{Assumptions}
\label{ass}

The simulations have been performed assuming a time step of $\Delta t=0.1~Gyr$.
This quantity has been chosen to be larger than the cluster crossing time, in
order to ensure that all the stars of a given mass group can approach the 
radial distribution predicted by their own King profile, and short enough to
avoid significant variations of the cluster structural parameters.
During each time step, the equivalent effect of the multiple interactions that 
occur for a binary population is calculated.
 
The width of the layers has been chosen to allow a higher resolution in the
inner (i.e. more populated) regions of the cluster.
I assumed $\Delta r=0.1~r_{c}$ for
$r<5~r_{c}$, $\Delta r=r_{c}$ for $5<r/r_{c}<50$ and $\Delta r=10~r_{c}$ for
$r>50~r_{c}$.

The cluster stars have been divided in mass bins of size $0.1~M_{\odot}$ in the 
mass range $0.1<M/M_{\odot}<5$. An additional bin formed by stars in the mass
range $5<M/M_{\odot}<120$ has been considered. These massive stars have
lifetimes shorter than the time-step of the simulation and are removed during 
the first stage of evolution (see Sect. \ref{code}). 
I adopted the Initial Mass Function (IMF) by Kroupa (2002), which can be written as a 
broken power law $dN=m^{\alpha}~dm$, with $\alpha=-1.3$ for
$0.1<M/M_{\odot}<0.5$, and $\alpha=-2.3$ for $M>0.5 M_{\odot}$.

The initial mass-ratios and periods distributions of the binary population have
been chosen in agreement with the results of Duquennoy \& Mayor (1991).
In particular, the initial mass-ratios distribution has been derived by random 
associating stars belonging to different mass bins.
I assumed a log-normal distribution of periods $g(log~P)$ centered at $log~P(d)=4.8$ and with
a dispersion $\sigma_{log~P}=2.3$, where P is the period expressed in days. 
The correspondig semi-axes distribution of
each group of binaries can be derived by considering the third Kepler's law
$$
a=\left[ \frac{G~(m_{1}+m_{2})~P^2}{4 \pi^2} \right]^{1/3} 
$$
The distribution of periods has been truncated assuming
an upper limit\footnote{I did not assumed a criterion to link the upper truncation of the period
distribution to the local velocity dispersion, assuming that binaries 
born in all clusters with the same initial properties.} of $P=10^7~d$ in 
agreement with Sills et al. (2003).
Moreover, a lower limit has been imposed by rejecting all the binaries
whose major semi-axes ($a$) turn out to be lower than their limiting orbital 
distance ($a_{min}$; see Sect. \ref{coal}).
The distribution of binary eccentricities has been assumed to follow a thermal 
distribution with probability density $p(e) = 2e$.

The mass-radius relation and the evolutionary timescales of each mass group 
have been derived from the models by Pietrinferni et al. (2006) assuming a
metal mass fraction $Z=10^{-3}$.

In order to limit the number of free parameters, I assumed 
the King parameter $W_{0}$ as a linear combination of the
central density and velocity dispersion
\begin{equation}
\label{eq_king}
W_{0}=\left[ log~\left( \frac{\rho_{0}}{M_{\odot}~pc^3} \right) -2.log~\left(
\frac{\sigma_{v,0}}{Km~s^{-1}}\right) \right]1.921+2.934
\end{equation}
This relation has been derived empirically using the central densities and 
velocity dispersions reported by Djorgovski (1993) and
the cluster concentrations by Trager, King \&  Djorgovski (1995).
In the same way, the core radius $r_{c}$ has been assumed to be linked to the
above parameters in the following way
$$r_{c}=\left( \frac{\sigma_{v,0}}{Km~s^{-1}}\right)  \left( \frac{M_{c}}{10^{2.67}
M_{\odot}} \right)^{-1/2} pc~~~\mbox{Bellazzini et al. (1998)}$$

The dependence of the obtained results on the choice of some of these assumptions is
investigated in Sect. \ref{test}.

\subsection{Tidal capture}
\label{tid}

A mechanism proposed to form binary systems is based on the tidal capture of a
companion induced by a close encounter between two single stars (Clark 1975).
Indeed, when two colliding stars approach with a relative velocity and 
an impact parameter lower than given limits, they can form a bound system.
The rate of binary systems formed through this mechanism has been evaluated
using the cross section $\sigma_{tc}$ provided by Kim \& Lee (1999).
I considered the collisions between two non-degenerate stars with a polytropic 
index $x=1.5$ (see their eq. 6).
In a time interval $\Delta t$, the number of tidal captured binaries having 
primary and secondary masses $m_{1}$ and $m_{2}$ and period $P$ turns out to be

$$
\Delta N_{b}(m_{1},m_{2},log~P)=
\int_{0}^{r_{rel}}\frac{d N_{b,tc}}{d V} d^3 r
$$
where
$$
\frac{d N_{b,tc}}{d V}=
\frac{n_{s,r}(m_{1})~n_{s,r}(m_{2})~\Delta
t~g(log~P)}{1+\delta_{1,2}}\times$$
$$
~~~~~~~~~~\int_{0}^{v_{e,r}} (1-F_{coll,v}) v~\sigma_{tc}~f(v,r)~d^3 v
$$

where $\delta_{1,2}$ is the Kronecker delta which is unity if stellar type 1 and 2
are identical and zero otherwise.
The quantity $F_{coll,v}$ is the fraction of collisions which lead to the
formation of an unstable binary system whose natural evolution is the
coalescence in a single massive star (see Sect. \ref{coal}) and can be written as
$$F_{coll,v}=\frac{\sigma_{coll}}{\sigma_{tc}}$$
As can be seen from the above equations, the binary systems formed via tidal capture
are assumed to follow the original period distribution.

\subsection{Collisional hardening and binaries ionization}
\label{ion}

When a single star collides with a binary systems, the two objects exchange part
of their energy. In particular, the colliding star gain (lose) a fraction of its
kinetic energy, which is balanced by an decrease (increase) of the binary system 
binding energy, if the $hardness~parameter$
\begin{equation}
\label{eq_eta}
\eta=\frac{E_{b}}{m_{s}~\sigma_{v,r}^2}\\
\mbox{where}\\
E_{b}=\frac{G~m_{1}~m_{2}}{2~a}
\end{equation}
    
is larger (smaller) than unity (Heggie 1975).
The variation of binding energy of the binary system translates into a change in 
the period (i.e. in the orbital separation).
This process is called "collisional hardening (softening)".
If the binding energy of the binary system drop to zero then the binary is
disrupted (binary ionization).

The code follows the evolution of the binding energy of each sub-population of
binaries and calculates the number of binaries which undergo ionization.
In particular, the fraction of binding energy gained (lost) by the binary system
after a time interval $\Delta t$ has been assumed to be 
$$
\frac{\Delta E_{b}}{E_{b}} \sim \Delta t~
\int_{0}^{r_{rel}}n_{b,r}(m_{1},m_{2},log P)
\left( \frac{1}{E_{b}} \frac{d E_{b}}{d t}\right)_{r} d^{3} r
$$
and
$$
\left( \frac{1}{E_{b}} \frac{d E_{b}}{d t} \right)_{r} =
\left\{
\begin{array}{rl}
\label{eq_ion}
\frac{7.6 \sqrt{\pi}~G~a~<n_{s}~m_{s}>}{\sigma_{v,r}} & \mbox{if } \eta>1\\
-\frac{16 \sqrt{\pi}~G~a~<n_{s}~m_{s}^2>~log \Lambda}{\sigma_{v,r} (m_{1}+m_{2})}
& \mbox{if } \eta<1\\
\end{array}
\right.
$$
(Binney \& Tremaine 1987; Hills 1992).\\
Here the quantities $<X>$ indicates the mean value of X measured at the radius
$r$. 
I assumed $\Lambda=1/\eta$, which is a suitable choice for
$\eta<1$ (Binney \& Tremaine 1987).
I neglected the change of the binary eccentricity distribution due to 
collisions.

Collisions between binaries plays also an important role, in particular when
high primordial binary fractions are considered.
Binary-binary interactions are very efficient for ionizing one (or both) of the two 
binaries (Mikkola 1983a; Sweatman 2007).
In a collision between binaries the most probable outcome is the disruption of the 
softer binary and hardening of the harder one.
As a first-order approximation, the collisions between two binary 
systems have been treated as normal
collisions between a binary and a single point-mass object with a mass equal to the systemic 
mass of the colliding binary system ($m_{1}+m_{2}$). 
Note that this simplification underestimates the amount of energy transferred
between the two binaries since the geometrical part of the cross sections
defined above would be few times larger then what assumed.  
Moreover, under this assumption only one of the various possible outcomes of a
binary-binary collision is considered (see Fregeau et al. 2004).
However, despite the adopted simplifications, the treatment described above 
accounts for most of the contribution of collisions between binaries to the 
ionization process.

Similarly, the variation of periods and semi-axes have been calculated considering that
\begin{equation}
\label{eq_a}
\frac{\Delta a}{a}=-\frac{\Delta E_{b}}{E_{b}}~~~~~~\mbox{and}~~~~~~
\Delta log~P=-\frac{2~ln~10}{3}\frac{\Delta E_{b}}{E_{b}}
\end{equation}

\subsection{Exchanges}
\label{exch}

A possible occurrence in a collision between a single star and a binary 
system consists in the formation of an unstable triple system which naturally
evolves by
expelling one of the components (usually the least massive) of the original 
system. 
The number of exchanges has been estimated by means of the cross section 
$\sigma_{exch}$ provided
by Heggie, Hut \& McMillan (1996) (see their eq. 4 and 17).
Since this formulation is valid only in the case $\eta>1$, I calculated the 
number of exchanges only in this regime. On the other hand, the process of
exchange in soft binaries is largely less efficient than the one of ionization 
and can be neglected. 
Therefore, the number of exchanges involving an hard binary having components of
masses $m_{1}$ and $m_{2}$ and period $P$ with a colliding star with mass $m_{s}$ 
is
$$
\Delta N_{exch}(m_{1},m_{2},m_{s},log~P)=\int_{0}^{r_{rel}}\frac{d N_{exch}}{d r} d^3 r
$$
with
$$
\frac{d N_{exch}}{d r}=
n_{s,r}(m_{s})~n_{b,r}(m_{1},m_{2},log~P)~\Delta t\times$$
$$
\int_{0}^{v_{e,r}} v~\sigma_{exch}~f(v,r)~d^3 v
$$
At each time-step, the number of binary sistems and single stars have been corrected for
this effect.
 
\subsection{Stellar evolution}
\label{stev}

As time passes, most stars evolve and die after timescales which depend mainly
on the star's mass. 
Massive stars ($M>7~M_{\odot}$) at the end of their evolution explodes as SNII.
Less massive stars suffers strong mass losses during their latest stages of
evolution and terminate their evolution as white dwarf. 

When a star belonging to a binary system evolves there are many possible outcomes:
If the mass of the primary star is large enough to produce a SNII explosion part
of the elapsed energy is converted in a velocity kick that increase the total
energy of the system.
If the total energy (potential and kinetic) is positive then the system is 
disrupted and its components will evolve separately. Else, the system remains in
a bound state in which one of the two companions is the compact remnant of the
evolved star.  

Otherwise, if the primary component of the binary system is small enough to
avoid the ignition of the triple-$\alpha$ cycle, mass losses produce a rapid
halting of the nuclear reactions leading to the formation of a white dwarf (WD).
Given the large number of low-mass stars, this event is the most probable among
the various possible outcomes.

In particular, I considered two important cases:
{\it i)} the envelope of the evolving stars reach the Roche-lobe of the companion,
igniting the process of mass-transfer (see Sect. \ref{coal}); {\it ii)} the
star follows an unperturbed evolution leading to the formation of a binary
system formed by a WD and a companion. 

To account for these processes, at each time-step of the simulation, 
a limiting mass has been associated. Then,  
For each mass bin, the code calculates the fraction of stars with masses
larger then this limiting mass, assuming the stars to populate uniformly 
the mass bin. When this process take place in a binary system, the occurrence of
the various possible processes described above has been considered.
In particular, in the case of a massive primary component I followed the
prescriptions of Belczynski et al. (2002).
In the case of binary systems formed by low-mass stars a distinction between
processes {\it i)} and {\it ii)} has been made using the criterion described in 
Sect. \ref{coal}.
The amount of mass lost at the end of the evolution has been calculated using
the prescriptions by Weidemann (2000).

The number of binary and single stars is updated accordingly.

\subsection{Direct collisions and Coalescence}
\label{coal}

From the early '50 it is well known that in globular clusters exists a 
population of massive objects that, in the color-magnitude diagram, lie along an extension 
of the Main Sequence, in a region which is brighter and bluer than the turn-off
(Piotto et al. 2004 and references therein). 
These stars are called Blue Straggler Stars (BSS).
Two mechanisms have been proposed to explain the origin of these stars: {\it I)}
direct collision between two single stars (Hills \& Day
1976) and {\it II)} mass-transfer activity
in close binary systems (McCrea 1964).
 
Case {\it I)} occurrs if the impact parameter of the collision between two 
single stars is small enough to form an unstable system in 
which the process of mass-transfer leads to the formation of a single massive 
star. 

In case {\it II)} when the orbital separation of a binary system 
reduces to a critical distance, the binary system became unstable and the 
ignition of the process of stable mass-transfer occurs.
This phenomenon can be induced by two different mechanisms: {\it IIa)} collisional 
hardening (see Sect. \ref{ion}) and {\it IIb)} off-Main Sequence evolution of the 
primary component of the binary system. 
In the former case, hard binaries subject to a high number of collisions tend to
reduce their orbital separation (see eqn. \ref{eq_a}). In the latter case, when the primary component
of the binary system exhausts the hydrogen in its core, it expands its envelope by 
a large factor (which
depends on its mass) possibly reaching the Roche-radius of the secondary component.

The critical distance between two companions in a binary system to ignite
stable mass-transfer is 
$$ a_{min}=2.17 \left( \frac{m_{1}+m_{2}}{m_{2}} \right)^{1/3}~R_{2}~~~\mbox{Lee \& Nelson (1988)}$$

The final product of these processes is a system containing a massive star having a mass
$m_{1}<m<(m_{1}+m_{2})$.  
The formation of BSSs affects the number of binary and single stars.
 
To evaluate the frequency of the processes described above, I first consider the
cross-section $\sigma_{coll}$ for a stellar collision which leads to a close binary system with
orbital separation $a<a_{min}$ (case {\it I}). 
This cross-section has been calculated using the 
formulation by Lee \& Ostriker (1986; see their eqns. 2.8 and 4.12)
$$ \sigma_{coll}=\pi a_{min} \frac{2 G m_{2}}{v^2}$$

Then, the number of coalesced stars via direct collision can be written as 
$$
\Delta N_{b}(m_{1},m_{2},log~P)=
\int_{0}^{r_{rel}}\frac{d N_{b,coll}(m_{1},m_{2})}{d r} d^3 r
$$
where
$$
\frac{d N_{b,coll}}{d r}=
\frac{n_{s,r}(m_{1})~n_{s,r}(m_{2})~\Delta t}{1+\delta_{1,2}}
\int_{0}^{v_{e,r}} v~\sigma_{coll}~f(v,r)~d^3 v
$$

The fraction of coalesced stars via collisional hardening (case {\it IIa}) has 
been calculated by following the evolution of the semi-axes of the binary
population (see Sect. \ref{ion}). 
Instead, when the coalescence is due to the evolution of the primary star of a
binary system (case {\it IIb}), the code calculates the fraction of system with an evolving
primary star which satisfies the condition
$$ a<a_{min} \frac{R_{1,max}}{R_{1}}$$
Where $R_{1}$ and $R_{1,max}$ are the radii of the primary star during its
quiescient and evoluted stages, respectively. 

\subsection{Evaporation}
\label{evap}

During its evolution, a globular cluster undertakes the tidal stress of its
host galaxy. As a result of this interaction, it lose part of its stellar systems
decreasing its total mass. 
The rate of evaporating systems depends on the cluster orbit and on its
structural parameters (Gnedin \& Ostriker 1999). 

Here, the treatment of evaporation has been greatly simplified assuming 
the cluster to lose a constant
fraction $\nu$ of its objects during all its evolution.
Although this is a crude approximation, it can be used to simulate the effect of
a tidal field on the relative amount of single and binariy stars survived to the
evaporation process.
A homogeneous evaporation rate is also in agreement with the most recent N-body 
simulation (Hurley et al. 2007; Kim et al. 2008).
At a given distance from the cluster center, I assumed the velocity distribution of a
population of objects with mass $m$ can be well represented by a Maxwellian
distribution\footnote{Although this assumption holds only in the core, I
extended its validity to the entire cluster only to calculate the relative 
fraction of evaporating stars belonging to different mass groups.} 
$$f(v,r)=A~e^{-\frac{v^2}{2 K}}$$
where $K$ is a term proportional to $\sigma^{2}$ and A is a
normalization factor.
The quantity $K$ depends on the stellar mass according to the status of
relaxation of the cluster (see Appendix A).  

The relative number of evaporating systems belonging to a given group
of mass can be therefore calculated by integrating over the entire cluster
extension the fraction of stars with a velocity higher than the escape velocity
($v_{e,r}=\sqrt{-2 \Phi_{r}}$). 
$$
N'(m)=B~\int_{0}^{r_{t}} n_{r}(m) \frac{d \nu'(m)}{d r} d^3 r
$$
where
$$
\frac{d \nu'(m)}{d r}=\int_{\beta
W_{r}(m)}^{+\infty}y^{1/2}~e^{-y} d y
$$
$$
W_{r}(m)=-\frac{\Phi_{r}}{\sigma_{v,0}^2}\\
$$
The term $\beta$ accounts for the state of partial relaxation (see Appendix A)
$$
\beta=\frac{m}{[\gamma<m>+(1-\gamma)m]}
$$
where
$$\gamma=\frac{N_{m}(r<r_{rel})}{N_{m}}$$
is the fraction of stars with mass $m$ 
located in the region where relaxation already occurred.

Once this ratios are calculated for all the mass groups,
the normalization factor $B$ can be derived by
imposing that 
$$\Delta N_{sys}=\sum_{i}N'(m_{i})=\nu N_{sys}$$

\subsection{Cluster dynamical evolution}
\label{clust}

During its evolution, the cluster core lose part of its mass as a result of 
the processes of mass-segregation and evaporation. To mantain its virial
equilibrium the cluster collapses increasing its central density and
velocity dispersion. A significant population of hard binaries feeds kinetic
energy into the cluster through binary-single and binary-binary interactions
which counteract the effects of evaporation.
This process can halt and even reverse the contraction of the core (Hills 1975;
Gao et al. 1991). 
Fregeau et al. (2003) showed that also a small fraction of hard binaries is
sufficient to support the core against collapse significantly beyond the normal 
core-collapse time.

I considered the evolution of the structural properties of the cluster in a
highly simplified way. In particular, the code
calculates at each time-step of the simulation the number of stars contained
in the cluster core and updates 
the core radius and the central velocity dispersion using the relations
$$
\frac{r_{c}'}{r_{c}}=\frac{E}{E'}\left( \frac{N'}{N} \right)^{2} ~\mbox{and}~~~\\
$$
$$
\frac{\sigma_{v,0}'}{\sigma_{v,0}}=\left( \frac{E'}{E}\frac{N}{N'}\right)^{\frac{1}{2}}
$$  
where
$$
\frac{E'}{E}=\left[\frac{ln(N'/2)}{ln(N/2)}\right]^{88~z(\xi)}
$$
Hills (1975)\\
Where the symbols X and X' indicate the values of X in two subsequent time-steps.
The term $E'/E$ represents the ratio between the cluster binding
energies in two subsequent time-steps due to the contribution of
binaries.
Here, all quantities refers to the cluster core and $\xi$ refers to the fraction
of hard binaries.
The exponent $z(\xi)$ is related to the heating of binaries due to binary-single
stars and binary-binary interaction and can be written as  
$$z(\xi)= (\xi/1.3)+10.3~\xi^{2}(1.2-1/<\eta>)~\mbox{Mikkola (1983b)}$$
Moreover, the King parameter has been updated using eq. \ref{eq_king}.
In this way, the cluster structural parameters satisfy the virial theorem 
during all the simulation.

In Fig. \ref{radius} the evolution of the core radius
for the models s5p3e5f50 and s5p3e5f10 are shown. 
As can be seen, the general trend of the
core radus evolution is characterized by a progressive 
contraction of the core induced by the ever continuing losses of stars.
In model s5p3e5f50 the heat generated by hard binaries halt the contraction and 
lead to a stabilization toward an equilibrium value.
The behaviour of the evolution of the cluster characteristic radii qualitatively
agree quite well with the most recent N-body simulation computations (see
Fregeau et al. 2003; Kim et al. 2008). The details of the core oscillations 
and post-collapse evolution are not reproduced by
the code as a consequence of the simplified treatment described above.
However, this should have only a minor effect on the predicted evolution of the
properties of the binary population.  
 
\begin{figure*}
 \includegraphics[width=12.cm]{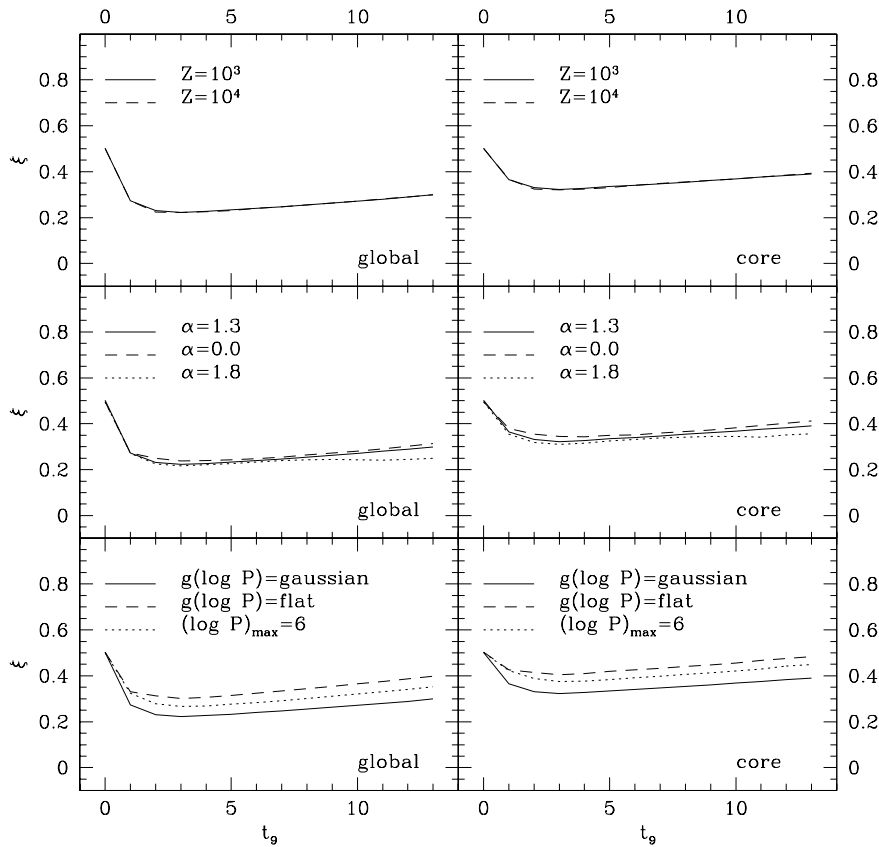}
\caption{The evolution of the binary fraction is shown as a function of time for
the reference model s5p3e5f50 (solid lines) and for the models calculated under different
assumptions. $Left~panels$ show the binary evolution in the whole cluster,
$right~panels$ show the binary evolution in the cluster core.}
\label{assum}
\end{figure*}
 
\begin{figure}
 \includegraphics[width=8.7cm]{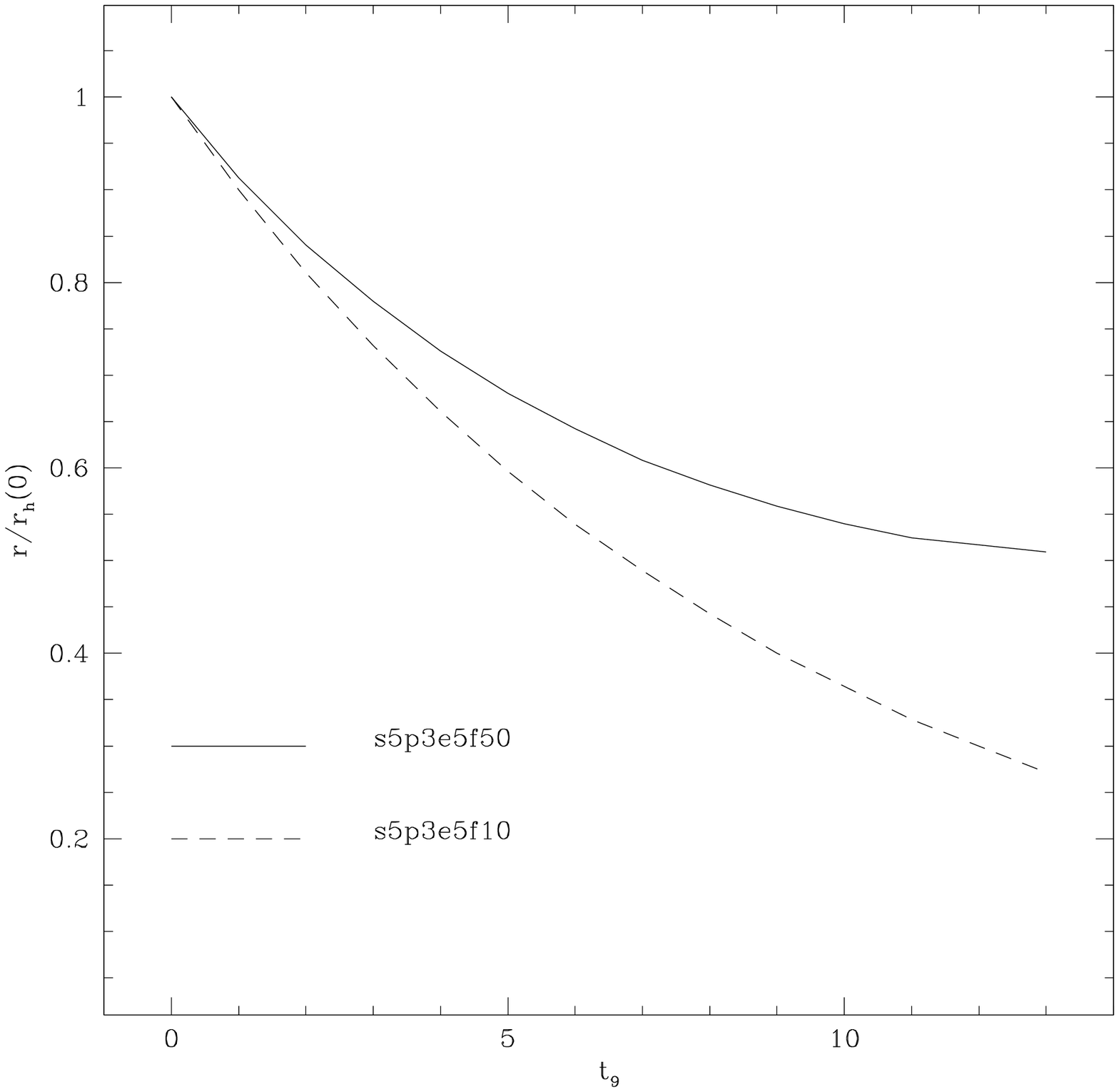}
\caption{The evolution of the core radius is shown as a function of time for
the models s5p3e5f50 (solid line) and s5p3e5f10 (dashed line).}
\label{radius}
\end{figure}

\section{Results and Dependence on the assumptions}
\label{test}
 
I run a set of 36 simulations spanning a wide range in density, velocity dispersion,
evaporation efficiency and initial binary fraction.
A list of all the performed
simulations and their initial and final parameters is provided
in Table 1. 

To test the dependence of the obtained results on the adopted assumptions in the 
following sub-sections I compare the model s5p3e5f50 with a set of simulations 
preformed with the same initial conditions but different assumptions.
The results of these additional simulations are listed in Table 2.

\begin{table*} 
 \centering
  \caption{Complete list of simulations.} 
\label{t:1}
  \begin{tabular}{@{}l|cccc|cr@{}}
  \hline
  \multicolumn{1}{l}{} &
  \multicolumn{4}{|c|}{initial conditions} &
  \multicolumn{2}{c}{final conditions}\\
  \hline
Model ID  & log~$\rho_{0}$     & $\sigma_{v,0}$ & $\xi$ & $\nu$         &  $\xi$ & $\xi_{c}$\\
          & $M_{\odot}pc^{-3}$ & $Km~s^{-1}$    & \%    & \% $Gyr^{-1}$ &  \%    &    \%    \\ 
 \hline
s2p1e0f10 & 1 & 2 & 10 & 0 &  8.3 & 12.4\\
s3p2e0f10 & 2 & 3 & 10 & 0 &  6.6 & 10.7\\
s5p2e0f10 & 2 & 5 & 10 & 0 &  6.6 & 10.0\\
s5p3e0f10 & 3 & 5 & 10 & 0 &  4.7 & 7.8\\
s5p4e0f10 & 4 & 5 & 10 & 0 &  2.8 & 5.3\\
s5p5e0f10 & 5 & 5 & 10 & 0 &  2.1 & 4.3\\
s9p4e0f10 & 4 & 9 & 10 & 0 &  2.9 & 5.0\\
s9p5e0f10 & 5 & 9 & 10 & 0 &  1.7 & 3.3\\
s13p5e0f10 & 5 & 13 & 10 & 0 & 1.6 & 2.9\\
s2p1e5f10 & 1 & 2 & 10 & 5 &  10.3 & 14.7\\
s3p2e5f10 & 2 & 3 & 10 & 5 &  7.5 & 11.0\\
s5p2e5f10 & 2 & 5 & 10 & 5 &  6.8 & 9.9\\
s5p3e5f10 & 3 & 5 & 10 & 5 &  4.8 & 7.6\\
s5p4e5f10 & 4 & 5 & 10 & 5 &  4.1 & 6.9\\
s5p5e5f10 & 5 & 5 & 10 & 5 &  3.1 & 5.7\\
s9p4e5f10 & 4 & 9 & 10 & 5 &  3.0 & 4.9\\
s9p5e5f10 & 5 & 9 & 10 & 5 &  2.5 & 4.1\\
s13p5e5f10 & 5 & 13 & 10 & 5 & 2.3 & 3.8\\
s2p1e0f50 & 1 & 2 & 50 & 0 &  40.9 & 51.5\\
s3p2e0f50 & 2 & 3 & 50 & 0 &  30.6 & 42.5\\
s5p2e0f50 & 2 & 5 & 50 & 0 &  31.5 & 41.8\\
s5p3e0f50 & 3 & 5 & 50 & 0 &  20.0 & 30.2\\
s5p4e0f50 & 4 & 5 & 50 & 0 &  12.1 & 21.0\\
s5p5e0f50 & 5 & 5 & 50 & 0 &  9.3  & 17.4\\
s9p4e0f50 & 4 & 9 & 50 & 0 &  11.6 & 18.8\\
s9p5e0f50 & 5 & 9 & 50 & 0 &  6.6  & 12.3\\
s13p5e0f50 & 5 & 13 & 50 & 0 &  6.0  & 10.5\\
s2p1e5f50 & 1 & 2 & 50 & 5 &  56.0 & 64.3\\
s3p2e5f50 & 2 & 3 & 50 & 5 &  44.6 & 53.8\\
s5p2e5f50 & 2 & 5 & 50 & 5 &  28.0 & 37.2\\
s5p3e5f50 & 3 & 5 & 50 & 5 &  29.9 & 39.0\\
s5p4e5f50 & 4 & 5 & 50 & 5 &  16.6 & 25.0\\
s5p5e5f50 & 5 & 5 & 50 & 5 &  11.5 & 20.2\\
s9p4e5f50 & 4 & 9 & 50 & 5 &  14.3 & 22.0 \\
s9p5e5f50 & 5 & 9 & 50 & 5 &  8.2 & 13.5\\
s13p5e5f50 & 5 & 13 & 50 & 5 & 7.9 & 12.7\\
\hline
\end{tabular} 
\end{table*} 

\begin{table*} 
 \centering
  \caption{Results of the test simulations.} 
\label{t:2}
  \begin{tabular}{@{}l|ccccc|cr@{}}
  \hline
  \multicolumn{1}{l}{} &
  \multicolumn{5}{|c|}{initial conditions} &
  \multicolumn{2}{c}{final conditions}\\
  \hline
Model ID  & $\xi$ & $log~Z$ & $\alpha (M<0.5 M_{\odot})$ & $g(log~P)$ & $log~P_{max}$ & $\xi$ & $\xi_{c}$\\
          &  \%   &         &                            &            & $d$           &  \%   &     \%  \\ 
 \hline
s5p3e5f50       & 50 & -3 & 1.3 & gaussian & 7 & 29.9 & 39.0\\
s5p3e5f50.zlow  & 50 & -4 & 1.3 & gaussian & 7 & 30.0 & 39.3\\
s5p3e5f50.m18   & 50 & -3 & 1.8 & gaussian & 7 & 24.9 & 35.7\\
s5p3e5f50.m0    & 50 & -3 & 0.0 & gaussian & 7 & 31.3 & 41.2\\
s5p3e5f50.pflat & 50 & -3 & 1.3 & flat     & 7 & 39.8 & 48.3\\
s5p3e5f50.p6    & 50 & -3 & 1.3 & gaussian & 6 & 35.2 & 44.9\\
 \hline
\end{tabular} 
\end{table*}

\subsection{Dependence on Metallicity}

As described in Sect. \ref{ass}, the simulations have been performed adopting a
metal mass fraction $Z=10^{-3}$. Metallicity influences both the evolutionary
timescales and the mass-radius relation and can in principle affect the evolution
of the binary population. 

In the upper panels of Fig. \ref{assum} the behaviour of the binary
fraction as a function of time is showed for the model s5p3e5f50 and for a
simulation performed with $Z=10^{-4}$. As can be noted, the two models are
indistinguishable even when the core binary fraction is considered.
Also the final binary fraction of the metal-poor model is the consistent with 
the reference metal-rich one within 0.3\% (see Table 2).

I conclude that the metallicity plays a neglegible role in the evolution of the
binary fraction in the presented simulations.  
Note however that metallicity can play an important role in the primordial
fraction of binaries and in the formation of X-ray binaries (see Ivanova 2006).  
In the models presented here such effects can remain hidden because of the
adopted simplifications.

\subsection{Dependence on the IMF}

Another important ingredient of the simulations is the IMF. Indeed, it determines the 
relative fraction of each mass group and influences the distribution of mass 
ratios.

Although the power-law exponent of the IMF is well estabilished for stars with
masses $M>0.5M_{\odot}$, there are still large uncertainties in the shape of the 
low-mass end of the IMF. Moreover, some authors claim a dependence of the
power-law exponent of the low-mass end of the IMF on metallicity (McClure et al.
1986; Djorgovski, Piotto \& Capaccioli 1993).

In the central panels of Fig. \ref{assum} the model s5p3e5f50 is compared with two simulations performed
using two extreme values of the IMF power-law exponent in the mass range
$M<0.5 M_{\odot}$ ($\alpha=0$ and $\alpha=1.8$).
As can be seen, the three models show a very similar behaviour during all the
cluster evolution. The final binary fractions of the three models are also
compatible within $\sim 5$\% .

Therefore, although the IMF plays a fundamental role in the determinations of the
main characteristics of the binary population, it plays only a minor role in the
evolution of the binary fraction. 

\subsection{Dependence on the Period distribution}

One of the most important assumptions made in the simulations regards the 
distribution of periods of the binary population.
The shape and the extremes of this distribution determine the ratio of
soft-to-hard binaries (i.e. the fraction of binaries which can undergo
ionization).

To evaluate the impact of these assumptions on the evolution of the binary 
fraction
I compared the model s5p3e5f50 with two simulations performed assuming {\it i)} 
a flat period distribution truncated at $log~P(d)=7$ (where P is expressed in days) 
and {\it ii)} a log-normal 
distribution truncated at $log~P(d)=6$.
The results of this comparison are shown in the lower panels of Fig. \ref{assum}. Note that, while a
variation of the upper limit of the period distribution produces only small
changes ($\Delta \xi < 6$\%)
in the final binary fraction, the model with a flat period
distribution mantains a significantly higher fraction of binary systems during
all the cluster evolution. In particular, after 13 Gyr, this last model contains
$\sim 10$\% of binaries more than the reference model.
As pointed out above, this effect is due to the higher fraction of hard binaries 
resulting from the flat distribution of periods (60\% at the beginning of the
simulation) with respect to the log-normal distribution (43\%).

All the results presented in the following sections are based on simulations run
assuming a log-normal distribution of periods truncated at $log~P(d)=7$. Although 
the dependence of the properties of the binary population on the cluster 
structural and environmental parameters should not be influenced by this choice, 
it is important to bear in mind that the absolute results of the individual 
simulations are highly sensitive to this assumption.   
 
\begin{figure}
 \includegraphics[width=8.7cm]{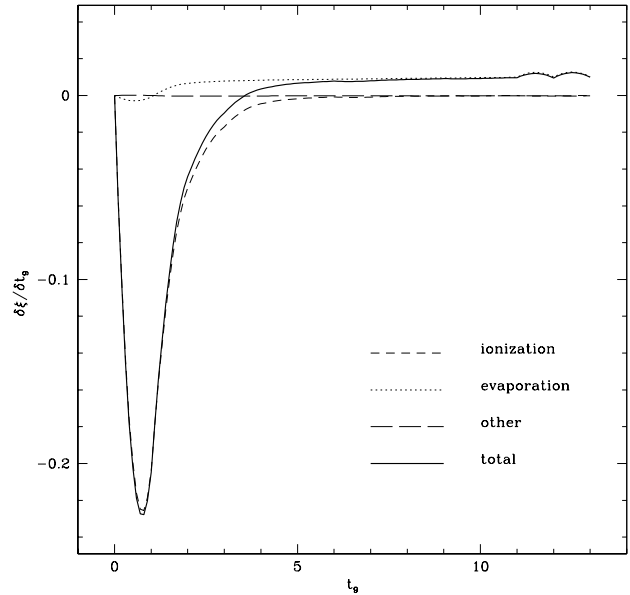}
\caption{Derivative of the binary fraction as a function of time for model
s5p3e5f50. The contributions 
of the various processes are indicated.}
\label{der}
\end{figure}

\begin{figure*}
 \includegraphics[width=12.cm]{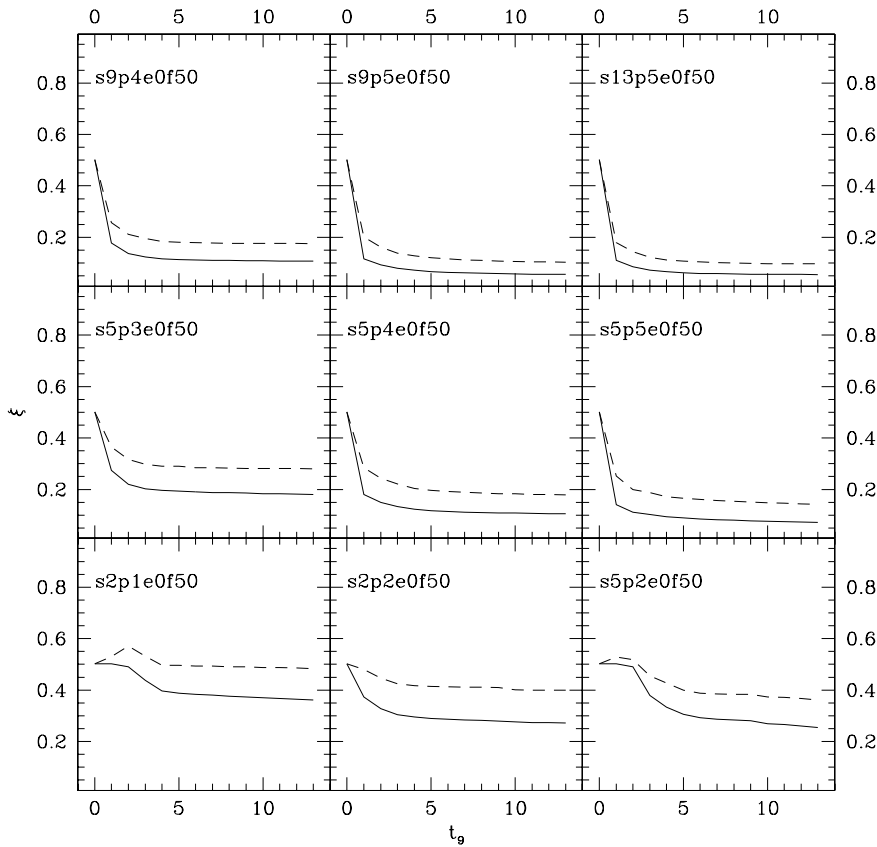}
\caption{The evolution of the binary fraction in the core (dashed lines) and in
the whole cluster (solid lines) are shown as a function of time for
all the models with $\nu=0$ and $\xi=50$\%.}
\label{all1}
\end{figure*}
  
\section{Evolution of the binary fraction}
\label{res}

The simulations listed in Table 1 cover a range of structural and
environmental parameters comparable to those spanned by Galactic globular
clusters.
Thus, they can be used to study the efficiencies of the various processes of
binary formation/destruction as a function of the initial conditions.

For this purpose consider the derivative
$$ \frac{d \xi}{d t_{9}}=\sum_{X} \left( \frac{d \xi}{d
t_{9}} \right)_{X}$$
This quantity can be seen as the sum of the contributions of the individual 
processes, where
$$\left( \frac{d \xi}{d t_{9}} \right)_{X}=\frac{N_{sys} d_{X}
N_{b}-N_{b} d_{X} N_{sys}}{N_{sys}^2} $$
represents the contribution of the process X to the above derivative.
In Fig. \ref{der} the behaviour of the derivatives defined above are shown as a function of
time for the model s5p3e5f50. Such a behaviour is qualitatively the same in the other
models with different initial conditions. 
As can be noted, the process of ionization is the main responsible for the 
variation of the binary fraction reaching its maximum
efficiency in the first Gyrs of evolution.
Another important contribution is given by evaporation 
(in models where $\nu \neq 0$) which enhances the fraction of binaries
sistematically during the whole cluster evolution. 
All the other processes play only a minor role.

\begin{figure*}
 \includegraphics[width=12.cm]{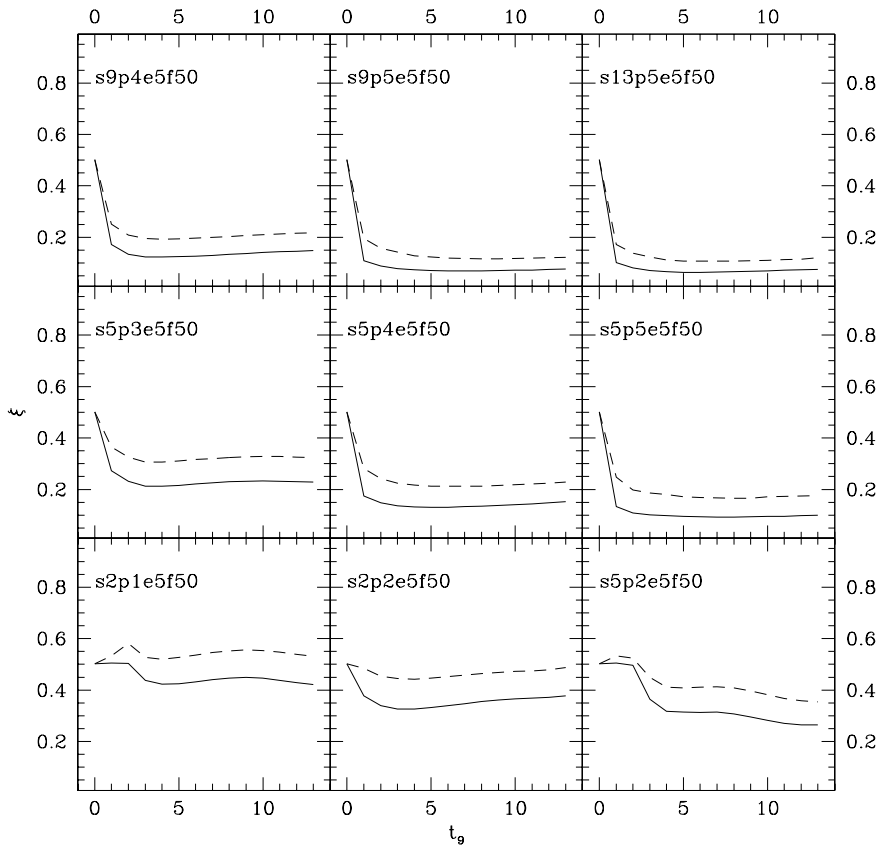}
\caption{Same of Fig. \ref{all1} but for
all the models with $\nu=5\% Gyr^{-1}$ and $\xi=50$\%.}
\label{all2}
\end{figure*}

In Fig. \ref{all1} and \ref{all2} the behaviour of the binary fraction as a function of time is
shown for all the simulations with initial binary fraction $\xi=50$\%.
As can be noted, the general evolution of the binary fraction is characterized by
a sudden decrease in the first Gyr, followed by a constant trend
during the subsequent Gyrs of evolution. 
In simulations run assuming $\nu \neq 0$, the fraction of binaries slightly increase in the 
last Gyrs of evolution.
The fraction of binaries in the core is sistematically larger than that of the
entire cluster. This difference is more evident after 2-3 Gyr evolution.

The strong initial decrease of the binary fraction is due to the process of 
binary ionization which destroy a
large fraction of binaries during the first Gyr of evolution. After this initial
stage, all the soft binaries have been destroyed and the evolution of the binary
fraction is driven by
the other processes which have a smaller efficiency. 
In particular, if the efficiency of the process of evaporation is large, the fraction of
binaries increases again.
Binary systems are in fact on average more massive then single stars and tends to have therefore smaller
velocities. Therefore, a larger
number of single stars reach the cluster escape velocity with respect to binary 
systems, thus increasing the relative fraction of binaries.
As relaxation proceeds a larger fraction of binaries sink into the central 
regions of the cluster, as a consequence of the process 
of mass segregation, thus increasing the core binary
fraction.   
 
\begin{figure*}
 \includegraphics[width=12.cm]{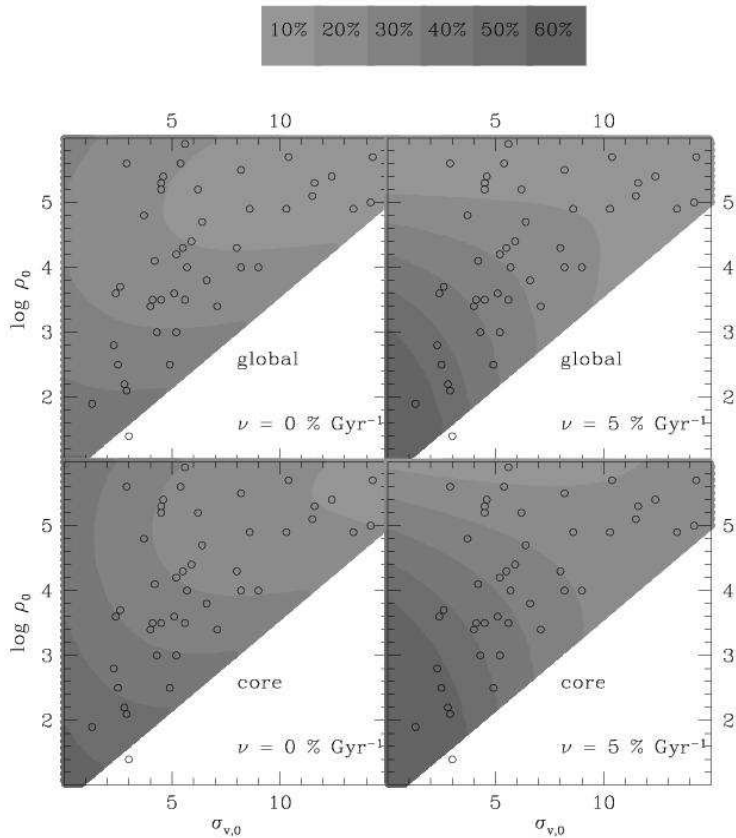}
\caption{Maps of the surviving binary fraction in the plane
$log~\rho_{0}-\sigma_{v,0}$ for $\xi=50$\%, $\nu=0$ ($left~panels$) and
$\nu=5\% Gyr^{-1}$ ($right~panels$). $Lower~panels$ refers to the core binary 
fraction, $upper~panels$ to the whole cluster binary fraction. 
Darker regions indicate higher binary fractions in steps of 10\%.
The locations of the Galactic globular clusters
(from Djorgovski 1993) are also marked with open circles.}
\label{map}
\end{figure*}

Fig. \ref{map} shows a map of the fraction of surviving binary systems after 13
Gyr in the $log~\rho_{0}-\sigma_{v,0}$ plane for the simulations with initial 
binary fraction $\xi=50$\%.
As can be seen, a clear trend is visible as a function of the cluster final
conditions. In particular:
\begin{itemize}
\item{The fraction of binaries decreases by increasing the central velocity
dispersion;}
\item{The fraction of binaries decreases by increasing the central density;}
\item{The fraction of binaries increases by increasing the efficiency of the
evaporation.}
\end{itemize}

The physical reasons at the basis of these trends are linked to the efficiencies 
of the processes of binary ionization and evaporation.
Indeed, by increasing the cluster velocity dispersion increase the ratio of
soft-to-hard binaries (see eq. \ref{eq_eta}) which are destroyed by the process
of ionization in the first
Gyrs of evolution. The efficiency of ionization increase also by increasing the
cluster density (see eq. \ref{eq_ion}). 
On the other hand, evaporation tends to increase the fraction of binaries (see above).  

\begin{figure}
 \includegraphics[width=8.7cm]{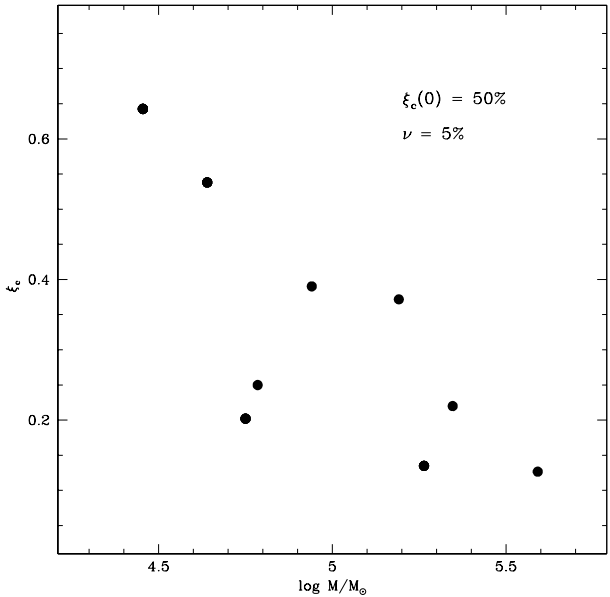}
\caption{The surviving fraction of binaries after 13 Gyr is shown as a
function of the residual cluster mass for all the models with $\xi=50$\% 
and $\nu=5$\%. A clear anticorrelation between these quantities is visible.}
\label{binm}
\end{figure}

Interestingly, the efficency of ionization and the cluster mass have the same 
dependences on the cluster structural parameters ($\rho_{0}$ and $\sigma_{v,0}$).
Indeed, more massive clusters have, on average, larger densities and 
velocity dispersions (Djorgovski \& Meylan 1993).
In Fig. \ref{binm} the fraction of binaries after 13 Gyr is plotted against the
residual cluster mass for all models with $\xi=50$\% and $\nu=5$\%. 
As expected, a clear anticorrelation is clearly visible between these two
quantities. The same result can be achieved considering different values of
the primordial binary fraction and evaporation efficency.

\begin{figure}
 \includegraphics[width=8.7cm]{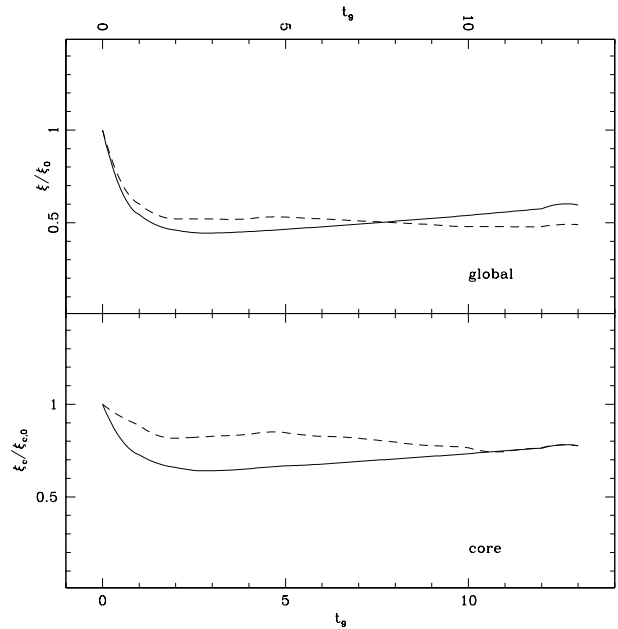}
\caption{The binary fraction normalized to the initial value is shown as a
function of time for the model s5p3e5f50 (solid lines) and s5p3e5f10 (dashed
lines). The $upper~panel$ refers to the entire cluster, the $lower~panel$ to the cluster core.}
\label{confb}
\end{figure}

In Fig. \ref{confb} the evolution of the binary fraction for the model s5p3e5f50
is compared with that calculated for a model with the same initial conditions
but a significatly smaller initial binary fraction (model s5p3e5f10). 
The binary fractions
for the two models have been normalized to their initial values in
order to compare the relative trend of their evolution.
Note that the model which starts with a lower binary fraction mantain a larger
fraction of binaries in the first stages of its evolution. 
This evidence is more evident when the core binary fraction is considered.
This is a consequence of two effects: {\it i)} the effect of the binary-binary
collisions and {\it ii)} the dependence of the local relaxation time on the mean
stellar mass (see eq. \ref{eq_tr}). 
Indeed, in clusters with higher binary fractions, binary-binary
collisions are $\sim$ 25 times more frequent and, given their higher energetic
budget, dominate the process of binary ionization.
Moreover, the mean stellar mass is smaller in clusters with smaller 
binary fractions. The local relaxation time in these clusters is therefore
longer, and a smaller fraction of binaries is therefore involved in the process
of binaries destruction. This effect is amplified in the core where more
binaries sink as a result of mass segregation.
The trend is reversed after few Gyrs when the impact of evaporation became more
important. In fact, models with higher fractions of binaries tend to mantain
binaries more efficently.
At the end of the simulation the two models contain a similar fraction of their
initial budget of binaries in the core regardless of their primordial content.    
In the following I discuss the dependence of the evolution of the binary
population considering the models calculated assuming an initial binary fraction
$\xi=50$\%. The conclusions are qualitatively the same for the models with a
smaller initial binary fraction. 

\section{Radial distribution of binaries}
\label{rad}

Binary systems are on average more massive then single stars. Therefore, they
are expected to have a
different radial distribution with respect to single stars, as a result of the 
process of mass segregation.

\begin{figure}
 \includegraphics[width=8.7cm]{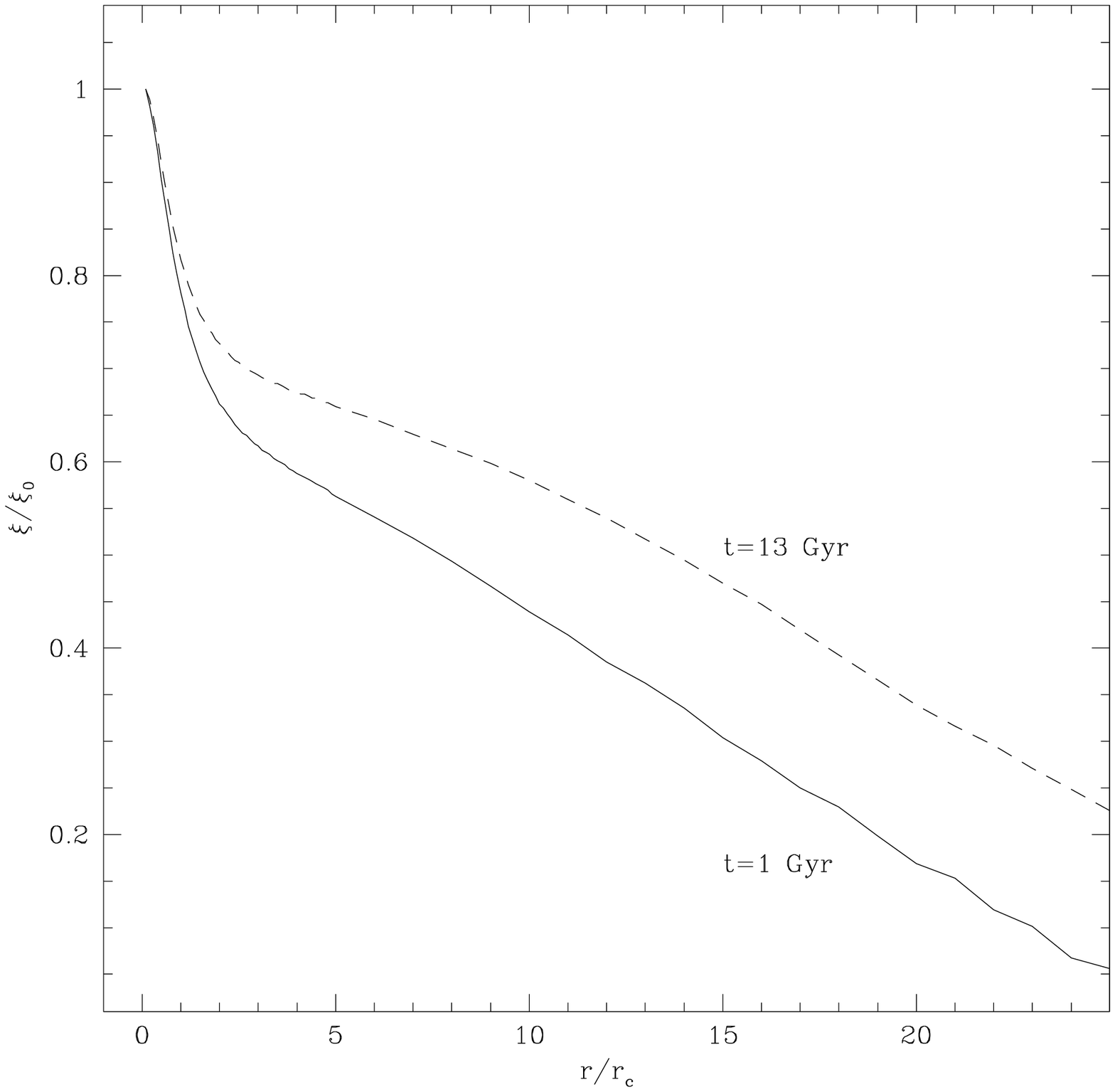}
\caption{Normalized binary fraction as a function of the radial distance to the
cluster center at different times from the beginning of the simulation for the
model s5p3e5f50.}
\label{radt}
\end{figure}

Fig. \ref{radt} shows the fraction of binaries (model s5p3e5f50) normalized to
the central value as a function of the 
radial distance from the cluster center at different times.
As expected, the fraction of binaries shows a central peak, decreasing toward the 
external regions of the cluster.
As time passes, relaxation occurs in the external regions and
the cluster increases its concentration as a result of the ever continuing 
losses of stars.
Consequently, the fraction of binaries populating the outermost part of the 
cluster slightly increase.

\begin{figure}
 \includegraphics[width=8.7cm]{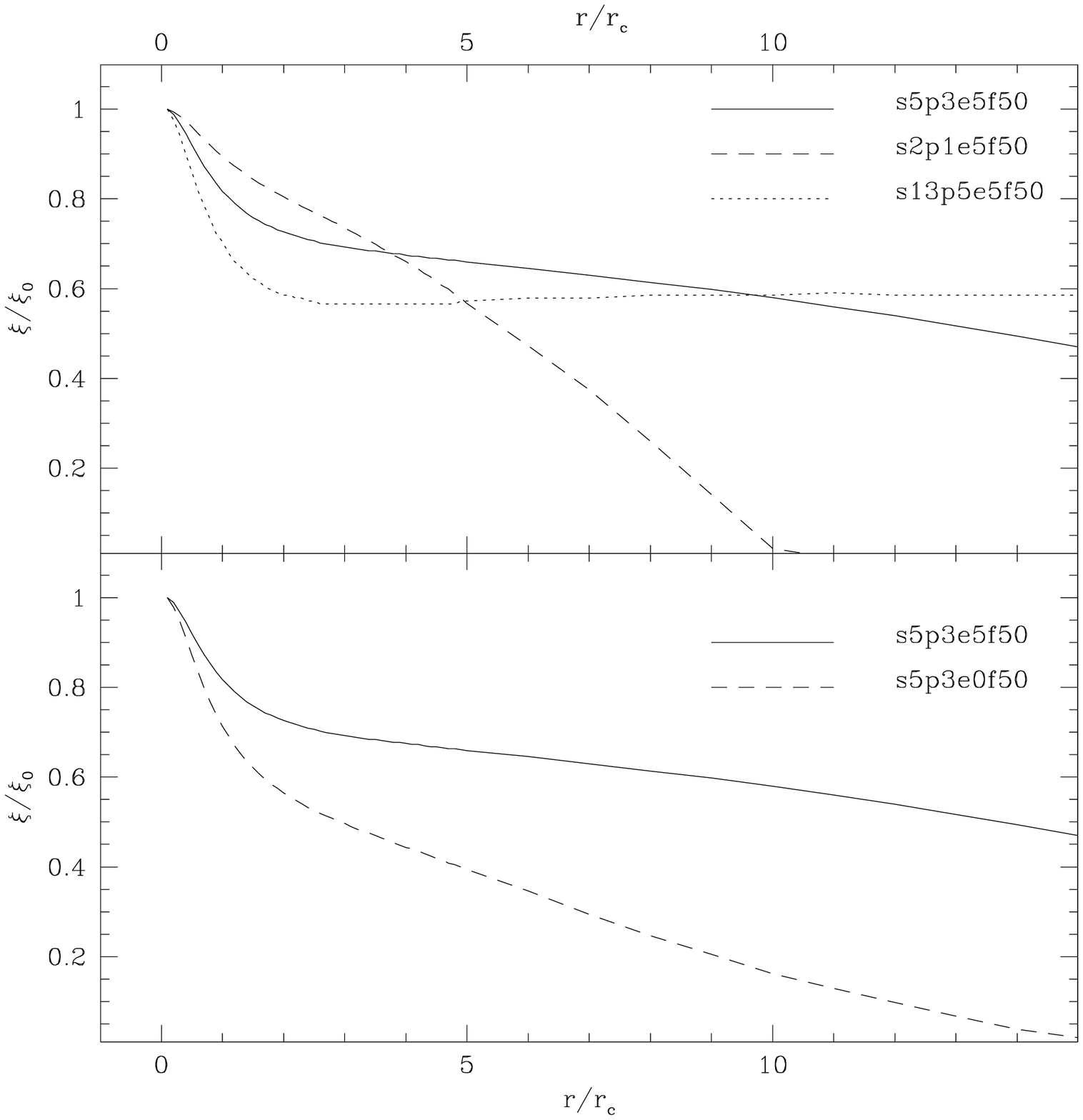}
\caption{In the $upper~panel$ the comparison among the normalized binary 
fractions as a function of the 
radial distance to the cluster center is shown for the model s5p3e5f50 (solid line),
s2p1e5f50 (dashed line) and s13p5e5f50 (dotted line).
The $lower~panel$ shows the same comparison but for the models s5p3e5f50 (solid line)
and s5p3e0f50 (solid line).}
\label{radtot}
\end{figure}

In Fig. \ref{radtot} the radial behaviour of the normalized fraction of 
binaries (model s5p3e5f50) calculated after 13 Gyr is compared with models with 
extremely different initial conditions (models s2p1e5f50 and s13p5e5f50; upper panel) and with a smaller
evaporation efficiency (s5p3e0f50; bottom panel).
Note that the slope of the decreasing trend of the binary fraction in the 
external part of the cluster is less steep in model s13p5e5f50 which undergo a
stronger dynamical evolution. The same behaviour is visible when evaporation 
accelerates the cluster dynamical evolution.  

Summarizing, the slope of the external decrease of the radial distribution of
the binary fraction is steeper as the cluster is dynamically younger.

A note of caution is worth: the code assumes that all the
cluster stars follow the radial distribution predicted by a given King model
according to their masses during the whole cluster evolution. 
Under this assumption, the process of mass segregation removes
instantaneously all the radial differences in binary fraction produced by the various
mechanisms of binary formation/destruction.
However, while this approximation is reasonable in the central regions (where 
the local relaxation time is relatively short) in the outer regions this
assumption could not hold. Therefore, the predicted radial distribution of 
binaries could not be reliable in the external regions of the cluster.

\section{Period Distribution}
\label{per}

As outlined in Sect. \ref{res}, one of the main processes that drives the evolution of
the binary fraction is the process of binary ionization. The binaries which are more subject to this
process are those with smaller binding energy (i.e. longer periods). The
shape of the distribution of periods therefore changes during the cluster evolution according to
the efficiency of the process of binaries ionization.

\begin{figure*}
 \includegraphics[height=7cm, width=15.cm]{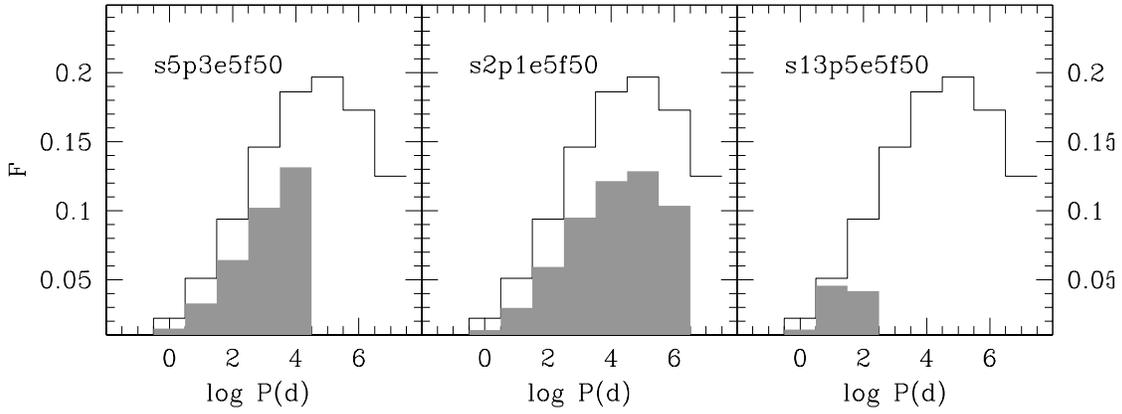}
\caption{Period distributions of three different models. Grey
histograms show the distribution after 13 Gyr. Open histograms show the
distribution at the beginning of the simulation. Periods are expressed in days.}
\label{lp}
\end{figure*}

In Fig. \ref{lp} the initial and final periods distributions of three models 
with different ionization efficencies are shown. 
As expected, in models with a high ionization efficency (see model
s13p5e5f50) all binaries with $log~P(d)>2$ are destroyed. On the oppsite side, when
ionization plays a minor role (model s2p1e5f50) a significant fraction of
binaries with period as long as $log~P(d)=6$ still survives. 

\section{Mass-ratio Distribution}
\label{fm}

During the cluster evolution also the mass-ratios distribution changes.
There are three main processes that drive its evolution: stellar
evolution, ionization and exchanges. Indeed, at the beginning of the simulation a large 
number of binaries are formed by a 
massive primary star which dies during the cluster evolution. The minimum
possible mass-ratio therefore increase with time.  
Moreover, the mean binding energy of the binary population is proportional to
the average mass-ratio. Thus, in clusters where ionization has a higher
efficiency the mass-ratios distribution should be more efficiently depleted up
to larger values of $q$.
During close encounters between a binary system and a massive colliding single star,
the secondary star of the binary system is usually ejected from the system.
When this process becomes frequent, the more stable systems against
exchanges are those formed by equal mass components. Thus, clusters with high
collisional rates should show an increase of the number of equal mass binaries. 

\begin{figure*}
 \includegraphics[height=7.5cm, width=12.cm]{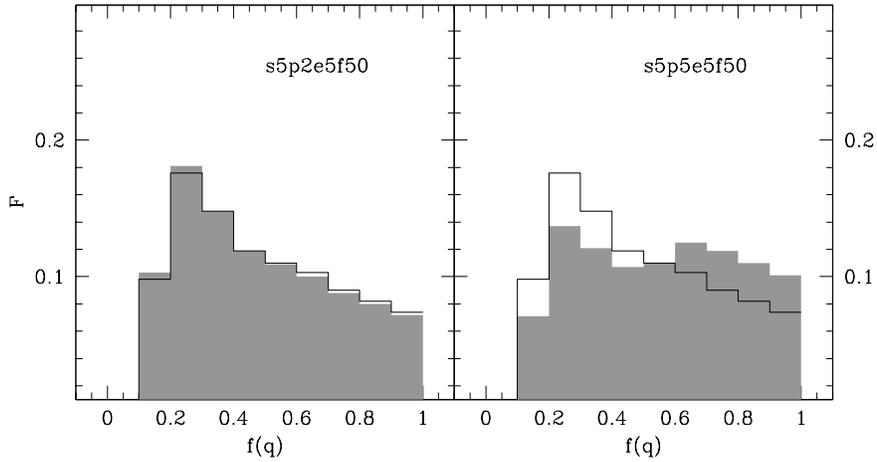}
\caption{The mass-ratios distributions after 13 Gyr of models 
s5p2e5f50 ($left~panel$) and s5p5e5f50 ($right~
panel$) are shown with grey histograms. 
The mass-ratios distribution of model s5p3e5f50 (open histograms) is 
shown in both $panels$ for comparison.}
\label{fq}
\end{figure*}

In Fig. \ref{fq} the distribution mass-ratios of the model s5p3f5e50 is compared
with those of two models with extremly different densities 
(models s5p2e5f50 and s5p5e5f50) and evaporation efficiency (model s5p3e0f50).
As expected, in all the distributions there are few systems with $q<0.2$, 
as a result of stellar evolution.
Moreover, the model s5p5e5f50 shows a lack of stars with low mass-ratios with
respect to the three other models. This is an effect of the high efficiency of
ionization which depletes the distribution of mass ratios up to larger values of $q$.
In none of the performed simulations an increase of the fraction of 
equal mass binaries is noticeable. This means that the efficiency of the
exchange process is rather low over the entire range of parameters spanned by
the simulations. Indeed, even in the densest systems, where collisions are more
frequent, the fraction of exchanges
$N_{exch}/N_{b}$ never exceeds few percent.
 
\section{Comparison with N-body and Monte Carlo simulations}
\label{confteo}

The analytical model presented here has been compared with the most recent
existing N-body (Hurley et al. 2007) and Monte Carlo (Ivanova et al. 2005)
simulations. 
As outlined in Sect. \ref{intro}, the two above approaches are completely different 
and lead to apparently discrepant results.
In fact, while Monte Carlo simulations by Ivanova et al. (2005) predict a strong
depletion of the core binary fraction, N-body simulations by 
Hurley et al. (2007) show the inverse trend.

Part of this discrepance can be due to the very different initial conditions of the
two simulations (Fregeau 2007). Indeed, Ivanova et al. (2005) simulated a series
of high
density cluster with a significant velocity dispersion. 
In the following I refer to their model B05 which has $n=10^5~pc^{-3}$ and
$\sigma_{v,0}=10~Km~s^{-1}$. In the approach followed by these authors, the
evolution of the binary population is supposed to proceed in a "fixed
backgroung" (i.e. a cluster whose structural parameters do not change during
the simulation). The fraction of evaporating stars in these simulation is also very
small ($\sim 6$\% after 13 Gyr; N. Ivanova, private communication).
This condition can be reproduced by assuming a very small efficency of
evaporation ($\nu\simeq0$).
Instead, Hurley et al. (2007) performed four different simulations assuming
lower densities and velocity dispersions for cluster which interact with the 
Galactic disk on a circular orbit in the solar neighborhood.
I considered their model K24-50 which have an initial binary fraction $\xi=50$\%
a central star density  $n=100~pc^{-3}$ and a velocity dispersion of 
$\sigma_{v,0}=3~Km~s^{-1}$. According to Fig. 4
of Hurley et al. (2007), the cluster lose $\sim70$\% of its object after 13 Gyr
(comparable with a model with $\nu=8\% 
Gyr^{-1}$).
Considering the results obtained in Sect. \ref{res} (see e.g. Fig. \ref{map})
the Hurley et al.'s model is expected to retain much more binaries with respect
to that of Ivanova et al. (2005). 

Moreover, the two above authors make different assumptions regarding the
distribution of periods. In fact, although both authors adopt a flat
distribution, Ivanova et al. (2005) assume an upper truncation 
at $log~P(d)=7$ while Hurley et al. (2007) impose an upper limit to
the orbital distance of 50 AU (corresponding to a truncation at $log~P(d)\sim5$,
see their Fig. 2). Consequently, the Ivanova et al.'s model contains a larger 
initial fraction of soft binaries which are easily destroyed during the cluster
evolution.

To test the consistency of the approach followed in this paper with the ones
quoted above, I run two simulations with the same assumptions and initial 
conditions of the above authors. I will refer to these two simulations as the
H-like and I-like models. Considering that both simulations start with an initial 
binary fraction
$\xi=50$\%, at the end of its evolution the H-like model contains a fraction of binaries
in the core of $\xi=76.9$\% while the I-like contains only a fraction of binaries
$\xi=25.1$\%. Although the absolute values of the above simulations differs 
from those reported by those authors, the qualitative behaviour of both
simulations is well reproduced. This means that {\it all the three approaches are
qualitatively self-consistent, and that the observed discrepancy ($\Delta
\xi\sim 52$\%) is due to 
both the intrinsic different starting parameters and the different assumptions}.   

To quantify the effect of the different assumptions on the resulting binary
fraction, I performed a series of
simulations adopting the starting conditions of the H-like and I-like models but
different values of the maximum period and evaporation 
efficiency. The complete set of simulations are summarized in Table 3.
As can be seen, adopting the same assumptions, the difference between the fractions of survived binaries predicted by
the two models reduces by a factor of two ($\Delta \xi = 21 \div 27$\%).
Therefore, at least an half of the difference between the fractions of 
survived binaries predicted by the above models is due to the different 
assumptions. 
The remaining difference can be addressed to the different initial cluster 
conditions. 

\begin{table*} 
 \centering
  \caption{Results of the H-like and I-like simulations.} 
\label{t:3}
  \begin{tabular}{@{}l|ccccc|cr@{}}
  \hline
  \multicolumn{1}{l}{} &
  \multicolumn{5}{|c|}{initial conditions} &
  \multicolumn{2}{c}{final conditions}\\
  \hline
Model ID  & $\xi$ & $log~\rho_{0}$        & $\sigma_{v,0}$ & $\nu$         & $log~P_{max}$ & $\xi$ & $\xi_{c}$\\
          &  \%   &   $M_{\odot} pc^{-3}$ & $Km~s^{-1}$    & \% $Gyr^{-1}$ & $d$           &  \%   &  \%  \\ 
 \hline
H-like & 50 & 2.35 & 3  & 8 & 5 & 73.7 & 76.9\\
Hp7e8  & 50 & 2.35 & 3  & 8 & 7 & 56.3 & 61.7\\
Hp5e0  & 50 & 2.35 & 3  & 0 & 5 & 48.8 & 60.3\\
Hp7e0  & 50 & 2.35 & 3  & 0 & 7 & 33.8 & 45.8\\
I-like & 50 & 5.4 & 10 & 0 & 7 & 15.7 & 25.1\\
Ip5e0  & 50 & 5.4 & 10 & 0 & 5 & 21.6 & 33.2\\
 \hline
\end{tabular} 
\end{table*} 
 
\section{Comparison with Observations}
\label{confobs}

The code described here has been used to interpret the observational evidence
that young globular clusters contain a higher fraction of binaries with respect
to older ones (Sollima et al. 2007).
As outlined in Sect. \ref{res}, the fraction of binaries should not show 
significant variations in the last Gyr of evolution.
Therefore, this trend could be due to {\it i)} differences in the primordial binary 
fractions or {\it ii)} to a different evolution of the binary 
populations linked to the different structural and environmental parameters of
these two groups of clusters. 
Indeed, the group of young globular clusters of
the sample of Sollima et al. (2007) is characterized by an age of $t_{9}\sim$ 7 Gyr,
a mean central density $<log \rho_{0}>=0.65 ~\mbox{(where}~\rho_{0}~\mbox{is
expressed in} ~M_{\odot} pc^{-3}$) and a velocity 
dispersion $<\sigma_{v,0}>=1.3 Km~s^{-1}$ which are significantly different from 
those of the old group of that
sample ($<t_{9}>=11 Gyr$, $<log \rho_{0}>=2.15$ and $<\sigma_{v,0}>=3.3 Km~s^{-1}$). 
According to the results found in Sect. \ref{res}, the group of young clusters 
should mantain a larger fraction of binaries during their evolution.

To test if these differences can be responsible for the difference in
the observed binary fractions, I tried to reproduce the mean observed 
conditions of the two above groups of globular clusters assuming the same
initial binary fraction. 
It is worth of noticing that the fraction of binaries derived by Sollima et al.
(2007) is calculated in a restricted range of masses and mass-ratios.
To compare the results of the simulations with the observations, I compared the
$minimum~binary~fraction~\xi_{min}$ defined in Sollima et al. (2007) with the ratio 
between the number of binaries with a primary star in the mass range
$0.5<m_{1}/M_{\odot}<0.8$ and a mass ratio $q>0.5$, and the number of objects 
in the same mass range. 
Of course, the relation between this quantity and the global
binary fraction depends on the distribution of mass-ratios. Given the arbitrary
choice of this distribution (see Sect. \ref{ass}) the absolute value of the 
initial fraction of binaries derived here is not reliable and can be only used
for differential comparisons.
According to Sollima et al. (2007), the groups of old and young globular 
clusters have $minimum~binary~fractions$ $\xi_{min}=6$\% and $\xi_{min}=20$\%
respectively.
 
The model that better reproduces the fraction of binaries in the old group of
globular clusters after 11 Gyr is characterized by an initial binary fraction of
 $\xi=8$\%.
I found no combinations of the initial parameters able to reproduce the observed
fraction of binaries in the young group of globular clusters without assuming 
a larger initial binary fraction.
This indicates that, {\it although the dynamical status of the young group of
globular clusters favors the survival of a larger fraction of binaries, the most
of the observed difference have to be due to
primordial differences in the cluster binary content}.
     
\section{Conclusions}
\label{concl}

I presented a code designed to simulate the evolution of the properties of a 
binary population in a dinamically evolving star cluster.
A number of simulations spanning a wide range of structural and environmental
parameters have been run. 

In general, the fraction of binaries appears to decrease with time by a 
factor 1-5, depending on the initial cluster parameters. In particular, the
fraction of binaries quickly decreases in the first 
Gyrs of evolution, as a result of the high efficiency of the ionization process
in this initial stage.

The analysis of the contributions of each mechanism of formation and distruction
of binaries indicates that the main processes that drive the evolution of the binary fraction
in globular clusters are the processes of binary ionization and evaporation. 
This result seems to
be confirmed by the observational fact that open clusters and low-density
globular clusters contain more
binary systems than dense high-velocity dispersion globular clusters 
(Sollima et al. 2007). 
As a consequence, the fraction of survived binaries increases when the cluster
structural parameters support a lower efficiency of the process of ionization
and in clusters subject to strong evaporation.
Moreover, also the final distributions of periods and mass-ratios 
change as a function of the cluster structural parameters.
In particular, clusters with a higher efficiency of binary ionization tend to
have binaries with shorter periods and higher mass-ratios.
At present, the range in central density and velocity dispersion spanned by the
sample of globular clusters with homogeneous estimates of binary fraction is
very small and does not allow a proper comparison (see Sollima et al. 2007). 
However, preliminary results by Milone et al. (2008) based on the
analysis of a larger sample of globular clusters seems to confirm this trend 
. These authors found a significant anticorrelation
between the fraction of binaries and the cluster luminosity (i.e. mass).
Indeed, more massive clusters have, on average, larger densities and 
velocity dispersions (Djorgovski \& Meylan 1993).
Therefore, the observed anticorrelation could be due by the fact that the cluster 
mass and the efficency of binary ionization have 
the same dependence on the cluster structural parameters ($\rho_{0}$ and
$\sigma_{v,0}$).

The dependence of the obtained results on the assumptions has been tested.
The evolution of the binary fraction appears to be non-sensitive to the
assumed cluster metallicity and to the shape of the low-mass end of the IMF.
A significant dependence on the initial shape of the distribution of periods 
has been found. In particular, a difference of $\sim10$\%
has been found by switching from a log-normal to a flat distribution.
Therefore, the decreasing trend of the binary fraction with time could be even
inverted assuming a different distribution of periods at least in those clusters
where the efficiency of the ionization process is small.
Unfortunately, the shape of this distribution is still largely uncertain 
(Halbwachs et al. 2003). Studies performed on large samples of binaries in the 
Galactic field are not able to distinguish between the various proposed 
distributions, suffering strong observational biases (Abt 1983; Eggleton et
al. 1989; Duquennoy \& Major 1991; Halbwachs 2003). 
Moreover, detections of long-period binary systems (with $P>10^6 d$) are limited 
by the intrinsic impossibility to detect photometric and/or kinematical 
variability over such large timescales.
Given the importance of this parameter, an indication of the true shape of the 
periods distribution would be valuable.

The predicted radial distribution of binary systems shows a decreasing trend
with central peak and a rapid drop toward the external regions of the
cluster.
The reason at the basis of this behaviour is linked to the process of 
mass segregation which produces a concentration of the massive binary systems
in the central region of the cluster. 
The higher concentration of binary systems has been already observed by several
authors in different clusters (Yan \& Reid 1996; Rubenstein \& Bailyn 1997; Albrow et al. 2001; 
Bellazzini et al. 2002; Zhao \& Bailyn 2005; Sollima et al. 2007).

The results of the code presented here have been compared with the most recent
N-body and Monte Carlo simulations available in the literature.
Despite the many adopted simplifications, the predictions of the code appear to be qualitatively consistent with the 
results of the above approaches.  
I estimated that at least an half of the difference between the fractions of 
survived binaries predicted by the N-body simulations by Hurley et al. (2007) 
and the Monte Carlo simulations by Ivanova et
al. (2005) is due to the different assumptions made by these authors regarding
the upper end of the periods distribution and the treatment of evaporation. 
The remaining
difference can be addressed to the different initial cluster conditions
assumed by these authors (as already suggested by Fregeau 2007).  

The code presented here has been used to interpret the differences in the 
binary fractions measured in the sample of globular clusters presented by 
Sollima et al. (2007).
There is no combination of initial parameters able to reproduce the observed
fraction of binaries observed in group of clusters with different ages unless
assuming a different initial binary content.
The group of young globular clusters in the sample of Sollima et al. (2007) 
is formed by three clusters (namely Terzan
7, Palomar 12 and Arp 2). These clusters are also the most distant from the Sun 
and they are thought to belong to the Sagittarius Stream 
(Bellazzini et. al 2003). Thus, they might
be stellar systems with intrinsically different origins and
properties, whose initial conditions could significant differ from those of 
"genuine" Galactic globular clusters (see also Sollima et al. 2008).
Future studies addressed to the estimate of the binary fraction in other
clusters suspected to have an extragalactic origin will help to understand how
the environmental conditions influence the original content of binaries of 
globular clusters. 

\section*{acknowledgements} 

This research was supported by the Instituto de Astrofísica de Canarias.
I warmly thank Antonino Milone and Natasha Ivanova for providing their preliminary results. 
I also thank the anonymous referee for his helpful comments and suggestions.

\appendix
\section{Partially-relaxed multi-mass King models}

The density and velocity distribution profiles of globular clusters are well
represented by King models (King 1966).
These models assume that cluster stars located at a distance r from the cluster
center follow a distribution of velocities 
\begin{equation}
\label{eq_kingm}
f(\epsilon,J)=A e^{-\beta J^{2}}(e^{-\frac{\epsilon_{r}}{K}}-1)$$ 
\end{equation}
(Mitchie 1963; King 1966; Gunn \& Griffin 1979).\\ 
where K is a quantity proportional to $\sigma_{v}^2$ that accounts for the cluster relaxation, $\epsilon_{r}$ is the energy
for unity of mass ($\epsilon_{r}=\Phi_{r}+v^{2}/2$), $J$ is the angular momentum,
 and A is a normalization factor. The term exp(-$\beta J^{2}$) accounts for
 anisotropies in the distribution function (here I set this quantity equal to
 unity).
The shape of distribution of stars in these models is complely defined by the
King parameter 
$W_{0}=-\Phi_{0}/\sigma_{v}^2$ .
The density profile can be derived from the above distribution considering the 
Poisson equation and the relation
$$ \rho_{r}(m)=\int_{0}^{v_{e,r}} f(v,r)d^{3} v$$

After a timescale comparable to the local relaxation time, collisions become 
frequent and produce the equipartition of the kinetic energy ($K\propto m^{-1}$).
Under this condition, the distribution of velocities becames a function of the
mass. Otherwise, the distribution of velocities does not depend on mass
($K=const$) and the above formulation reduces to the case of a single-mass King 
model. 
However, relaxation occurs after different timescales at different distances
from the cluster center (see eq. \ref{eq_tr}). Therefore, only a fraction of
stars follow relaxed orbits.

For this reason, the code calculates at each time-step of the simulation the
fraction of stars located in the region where relaxation already occurred. 
Defining 
$$\gamma=\frac{N_{m}(r<r_{rel})}{N_{m}}$$
the distribution of velocities has been then
assumed to follow the form of eq. \ref{eq_kingm} where
$$
K=\frac{\gamma<m>+(1-\gamma)m}{m}\sigma_{v}^2
$$ 
The density and velocity dispersion profiles have been therefore calculated
accordingly.

\label{lastpage}

\end{document}